\documentclass[aps,prl,twocolumn,showpacs,groupedaddress]{revtex4} 

\usepackage{graphicx} 

\usepackage{dcolumn} 

\usepackage{bm} 

\usepackage{amssymb} 

\usepackage{t1enc}

\usepackage[danish]{babel}

\renewcommand{\vr}{\vek{r}} 
\newcommand{\vek}[1]{\mathbf{#1}} 
\newcommand{\nt}{\tilde{n}}

\begin{document}

\title{Ab-initio non-equilibrium quantum transport and forces with the real space projector augmented wave method}

\date{\today}

\author{Jingzhe Chen}
\author{Kristian S. Thygesen}
\author{Karsten W. Jacobsen}
\email{kwj@fysik.dtu.dk}
\affiliation{Center for Atomic-scale Materials Design, Department of Physics, Technical University of Denmark, DK-2800 Kgs. Lyngby, Denmark}
\begin{abstract}
  We present an efficient implemention of a non-equilibrium Green
  function (NEGF) method for self-consistent calculations of electron
  transport and forces in nanostructured materials. The electronic structure is
  described at the level of density functional theory (DFT) using the
  projector augmented wave method (PAW) to describe the ionic cores
  and an atomic orbital basis set for the valence electrons. External
  bias and gate voltages are treated in a self-consistent manner and
  the Poisson equation with appropriate boundary conditions is solved
  in real space. Contour integration of the Green function and
  parallelization over k-points and real space makes the code highly
  efficient and applicable to systems containing several hundreds of
  atoms. The method is applied to a number of different systems
  demonstrating the effects of bias and gate voltages, multiterminal
  setups, non-equilibrium forces, and spin transport.
\end{abstract}

\maketitle

\section{Introduction} 
Electron transport across nanostructured interfaces is important in a range 
of different areas including nano-electronics, organic photovoltaics,
and electrochemistry. First-principles modelling of electron transport
at the nano-scale has so far mostly been applied to molecular
junctions consisting of molecules contacted by metallic
electrodes\cite{smit02,djukic,venkataraman,quek,evers}.  However, more
recent applications also include graphene
nanoribbons\cite{sahin,novaes}, semiconducting and metallic nanowires\cite{markussen,lee,hirose}, and
bulk tunnelling junctions for magneto-resistance and electrochemical applications\cite{rungger,chen}. The rapid
developments in these areas towards atomic-scale control of interface structures, and the continuing miniaturization of
electronics components makes the development of efficient and flexible
computational tools for the description of charge transport at the
nano-scale an important endeavour.

The vast majority of first-principles electron transport studies have
been based on density functional theory (DFT) within the local
density (LDA) or generalized gradient (GGA) approximations. This approach is in
principle unjustified because the eigenvalues of the effective
Kohn-Sham Hamiltonian do not represent the true quasiparticle energy
levels. In particular, for tunneling junctions the energy gap between
the highest occupied states and lowest unoccupied states is too small\cite{ref3,juanma} and this 
can lead to an overestimation of the conductance. More accurate calculations incuding self-interaction corrections\cite{toher} and more recently the many-body GW approximation\cite{strange,strange2,rangel} yield conductance
values in better agreement with experiments. On the other hand, the NEGF-DFT approach often
provides a satisfactory qualitative description\cite{hybertsen,evers}
and its computational simplicity makes it a powerful tool for
addressing non-equilibrium properties of complex systems. It should be
mentioned that the formal problems associated with the use of DFT for
transport are overcome by time-dependent DFT (TDDFT) which allows for
an, in principle, exact description of the (longitudinal) current due
to an externally applied field\cite{stefanucci_almbladh}. However, it has been recently found that the
standard TDDFT exchange-correlation potentials do not yield any
improvement over the NEGF-DFT in terms of accuracy in predicting
conductance\cite{niehaus}.

In addition to the electronic current it is of interest to model the
forces acting on the atoms under non-equilibrium conditions, i.e.
under a finite bias voltage. Such forces ultimately determine the
stability of current carrying molecular
devices\cite{todorov,hedegaard}, but can also be exploited to
deliberately control the motion of single molecules by e.g. injecting
electrons into the molcular orbitals using a scanning tunneling
microscope (STM).

In this paper we describe the implementation of the NEGF-DFT method in
the GPAW\cite{GPAW_review,gpaw-lcao} electronic structure code. In GPAW the electronic states can be
described either on a real space grid or using an atomic orbital basis
set. For the NEGF calculations, the Green function is expanded in the
atomic orbital basis while the Poisson equation is solved in real
space. Contour integration and sparse matrix techniques together with
parallelization over both k-points and real space is exploited for
optimal efficiency. Although the basic elements of our implementation
are not new and have been described in earlier papers\cite{taylor,BrandbyNEGF,xue}, the 
possibility of applying a general gate and/or finite
bias voltage, the use of multiple leads, and inclusion of
non-equilibrium forces on the ions provides a flexible and
efficient computational platform for general purpose modelling of charge transport at the nano-scale and should be of interest to a large and growing community.

This paper is organized as follows. In Section 2 the transport model
and formalism are introduced. In Section 3 we describe the complex
contour integration technique used to obtain the non-equilibrium
electron density from the Green function. Section 4 describes the use
of sparse matrix methods, and in Section 5 we discuss the real space
solution of the Poisson equation. A number of illustrative
applications are presented in Section 7.

\section{Method} 
\begin{figure}
	[ht] \centering 
	\includegraphics[width=.5 \textwidth]{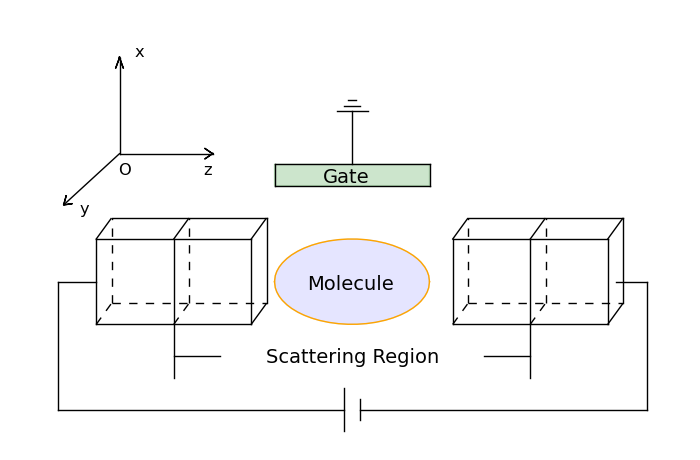}
        \caption{A scattering region including
          the nanostructure of interest (e.g. a molecule) and part of the electrode atoms is sandwiched
          between two semi-infinite electrodes. Periodic boundary
          conditions are used in the $x, y$ directions and open
          boundary conditions in the $z$ direction. The electron
          potential in the electrodes is periodic and can be obtained
          from a groundstate DFT calculation employing periodic
          boundary conditions in all directions. The Hartree potential
          at the scattering region boundary, which is used as boundary
          condition for the Poisson equation, is also obtained from
          the electrode calculation. The whole system can be subject
          to an external bias or gate voltage, and the electronic
          structure of the scattering region is calculated
          self-consistently in the presence of such external fields.}
	\label{fig:mainmodel}
\end{figure}

The transport model is shown in Fig.~\ref{fig:mainmodel}. Following the standard approach, the system is divided into left and right electrodes and the central scattering region (see the detailed description in the caption). The Hamiltonian of the system is given by (all notation related to PAW methodology is consistent with earlier GPAW papers\cite{GPAW_review, gpaw-lcao}) 
\begin{equation}
	\label{model:pseudo hamiltonian matrix} \tilde{\hat{H}} = -\frac{1}{2}\nabla^2 + \tilde{v}_{eff} + \sum_{aij}|\tilde{p}_i^a\rangle\Delta H_{ij}^a\langle\tilde{p}_i^a|, 
\end{equation}
where $a$ denote the atoms in the system and $i,j$ label the PAW projector functions of a given atom. Using a (non-orthogonal) atomic orbital basis set, the Hamiltonian can be written in the following generic form
\begin{equation}
	\label{model:scattering region hamiltonian} \left( 
	\begin{array}{ccc}
		H_{LL} & H_{LC} & 0 \\
		H_{CL} & H_{CC} & H_{CR} \\
		0 & H_{RC} & H_{RR},
	\end{array}
	\right) 
\end{equation}
The ``on-site'' Hamiltonian matrices of the electrodes, $H_{LL}$ (left) and $H_{RR}$ (right), and the coupling matrices $H_{LC}$ and $H_{RC}$, can be obtained from a homogeneous bulk calculation. If a bias voltage $V$ is applied, the matrices corresponding to the left and right electrodes should be shifted by $eV$ relative to each other, e.g. $H_{LC} \to H_{LC} +eVS_{LC}$ and $H_{LL} \to H_{LL} +eVS_{LL}$, where $S$ denotes the overlap matrix. We assume that there is no coupling between basis functions belonging to different electrodes. This assumption can be always satisfied by making the scattering region large enough.

The retarded Green function is written as
\begin{equation}
	\label{eq:retarded-green-function} G^r(\varepsilon) = [(\varepsilon)S-H_{CC}-\Sigma^r_L(\varepsilon) - \Sigma^r_R(\varepsilon)]^{-1} 
\end{equation}
The self-energies, $\Sigma^r_{L/R}$, represent the coupling to the electrodes and are obtained using the efficient decimation technique \cite{guinea}.

The lesser Green function is written as\cite{jauho} 
\begin{equation}
	\label{eq:keldysh-green-function} G^<(\varepsilon)=G^r(\varepsilon)\Sigma^<(\varepsilon)G^a(\varepsilon)+(1+G^r\Sigma^r)G_0^<(1+\Sigma^aG^a). 
\end{equation}
The latter term is nonzero for truly bound states and vanishes for states acquiring any width.

The pseudo density matrix (for the pseudo wave in the PAW framework) is the integral of $G^<$ 
\begin{eqnarray}
	\label{eq:density-matrix} D&=&\frac{1}{2\pi i}\int_{C}(G^r(z)-G^a(z))dz +\frac{i}{2\pi}\int_{E_f-eV/2}^{E_f+eV/2}G^{<}(\varepsilon^{\dagger})d\varepsilon \nonumber\\
	&&- 2\pi\theta i\sum_i G^r(\epsilon_i)
\end{eqnarray}
Here $\theta=k_BT$ and $C$ is a contour for the integral to be discussed more in Sec.~3.

The non-equilibrium electron density is obtained as 
\begin{equation}
	\tilde n(\mathbf{r})= \sum_{\nu \mu} D_{\nu \mu}\Phi_\nu (\mathbf{r})^* \Phi_\mu(\mathbf{r}) + \sum_a\tilde n_c^a, \label{eq:density}
\end{equation}
where $\Phi_{\nu}$ is an atomic orbital basis function and $\tilde n_c^a$ is the atomic pseudo core density. As is standard in the PAW formalism a tilde indicates a smooth quantity as opposed to an all electron quantity. The smooth charge density is given by
\begin{equation}
	\tilde{\rho}(\vr)= \nt(\vr) + \sum_a \sum_{\ell m} Q_{\ell m}^a \hat{g}_{\ell m}^a(\vr), 
\end{equation}
where $Q_{\ell m}^a$ are multipole moments and $\hat{g}_{\ell m}^a(\vr)$ is a so-called shape function. The last term is the contribution to the charge density coming from the positively charged nuclei. 

The effective potential is found as
\begin{equation}
	\tilde{v}= \tilde{v}_{coul}+\tilde{v}_{xc}+\sum_a\bar{v}^a, \label{effective_potential} 
\end{equation}
where the Coulomb potential satisfies the Poisson equation $\nabla^2\tilde{v}_{coul}=-4\pi\tilde{\rho}$, while $\tilde v_{xc}$ and $\bar{v}^a$ are the exchange-correlation potential and zero-potential, respectively. $\bar{v}^a$ is a parameter chosen to smoothen $\tilde{v}$ and which vanishes outside the augmentation sphere of atom $a$\cite{Kresse}.

To obtain self-consistency we thus have the iteration process $D\rightarrow \rho \rightarrow V_{eff} \rightarrow H \rightarrow D \rightarrow ...$. After convergence the current can be calculated by 
\begin{equation}
	I(V)=\frac{1}{\pi}\int_{-\infty}^{\infty}[f_L(\varepsilon)-f_R(\varepsilon)]\text{Tr}[\Gamma_L(\varepsilon)G^r(\varepsilon)\Gamma_R(\varepsilon)G(\varepsilon)^{\dagger}]\text{d}\varepsilon, 
\end{equation}
where $\Gamma_{L/R}(\varepsilon)=i(\Sigma_{L/R}^r(\varepsilon)-\Sigma_{L/R}^r(\varepsilon)^{\dagger})$.
For a derivation of the current formula we refer the reader to Ref.~\cite{meir} (orthogonal basis) or Ref.~\cite{nonorthogonal} (non-orthogonal basis).

The non-equilibrium force is obtained from the derivative of the total energy with respect to atomic positions. In the PAW framework, the total energy is written 
\begin{equation}
	E = \tilde{E} + \sum_a(E^a-\tilde{E}^a),
\end{equation}
with 
\begin{eqnarray*}
	\tilde{E} &=& \sum_{\nu \mu}\rho_{\nu \mu}\langle\Phi_{\mu}|-\frac{1}{2}\nabla^2|\Phi_{\nu}\rangle + \frac{1}{2}\int\frac{\tilde{\rho}(\vr)\tilde{\rho}(\vr)'}{|\vr - \vr'|}d\vr d\vr'\\
	&& + \sum_{a}\int\tilde{n}(\vr)\bar{v}^a(\vr)d\vr + E_{xc}[\tilde{n}] 
\end{eqnarray*}

The force can be obtained as
\begin{equation}
	\mathbf{F}^a=-\frac{
	\partial E}{
	\partial\mathbf{R}^a},
	\label{eq:force}
\end{equation}
where
\begin{eqnarray}
	\frac{
	\partial E}{
	\partial\mathbf{R}^a} &= &\sum_{\nu \mu}\frac{
	\partial E}{
	\partial\rho_{\nu \mu}}\frac{
	\partial\rho_{\nu\mu}}{
	\partial\mathbf{R}^a} + \sum_{\nu \mu}\frac{
	\partial E}{
	\partial T_{\nu \mu}}\frac{
	\partial T_{\nu\mu}}{
	\partial\mathbf{R}^a} \nonumber\\
	&&+ \sum_{L}\int\frac{\delta E}{\delta \tilde{g}^a_L(\vr)}\frac{d\tilde{g}^a_L(\vr)}{d\mathbf{R}^a}d\vr \nonumber\\
	&&+ \int\frac{\delta E}{\delta \tilde{n}(\vr)}\frac{
	\partial\tilde{n}(\vr)}{
	\partial\mathbf{R}^a}d\vr + \int\frac{\delta E}{\delta \bar{v}^a(\vr)}\frac{d\bar{v}^a(\vr)}{d\mathbf{R}^a}d\vr \nonumber\\
	&&+ \sum_{bij}\frac{
	\partial E}{
	\partial D_{ij}^b}\frac{
	\partial D_{ij}^b}{
	\partial\mathbf{R}^a}
	\label{eq:force2}
\end{eqnarray}
We note that the expression given above does not include the recently discussed Berry phase contributions to the nonequilibrium force\cite{hedegaard}.

\section{Numerical details} 
\subsection{Contour integration technique}

\begin{figure}
	[ht] \centering 
	\includegraphics[width=.5
	\textwidth]{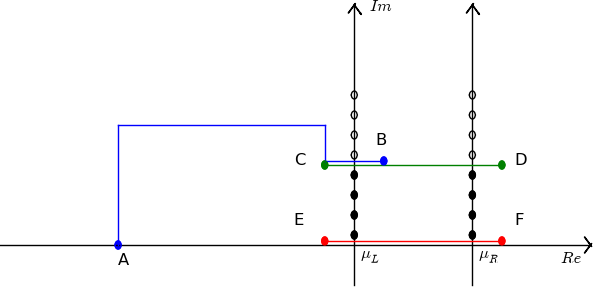} \caption{The contour used for the Green function integral in the complex energy plane. The coordinates of the indicated points are: A($E_{min}$, 0), where $E_{min}$ is less than all the eigen-energies of the system, which is usually taken as $\mu_L-100eV$ since we only calculate the valence electrons states; B($\mu_L+mk_BT, \Delta$), where $m$ satisfies $e^{-m}\approx0$(a typical value of $m$ is 10), $k_B$ is the Boltzmann constant, $T$ is the electron temperature, $\Delta$ is normally between $\epsilon_n$ and $\epsilon_{n+1}$, where $\epsilon_i=(2i-1)\pi$ (with $i$ a positive integer) is a pole of the Fermi-Dirac distribution function, thus the singulars below $\Delta$ should be counted when summing up the residues; C($\mu_L-mk_BT$, $\Delta$); D($\mu_R+mk_BT$, $\Delta$); E($\mu_L-mk_BT$, $\eta$), $\eta$ is a infinitesimal to avoid the inversion divergence; F($\mu_R+mk_BT$, $\eta$).} \label{fig:contour}
\end{figure}

The contour for the Green function integral in
Eq.~(\ref{eq:density-matrix}) is shown in Fig.~\ref{fig:contour}. The
retarded and lesser Green functions are integrated along the path AB
(see Fig.~\ref{fig:contour}) in the upper half plane and the path EF
closely above the real axis in the bias window\cite{BrandbyNEGF,
  TaylorNEGF}.

For the retarded Green function we use Gaussian quadrature by which a precision corresponding to a $2N-1$ order polynomial can be obtained by $N$ points.
We use an adaptive method to find the energy points necessary to obtain a sufficient precision\cite{LiruiThesis}: for a given region $[c-h, c+h]$, the integral $Q$ of a function $f$ can be estimated with the Gauss-Lobatto formula, 
\begin{equation}
	Q = \frac{h}{6}(f(c-h) + 5f(c-\frac{1}{\sqrt{5}}h) + 5f(c+\frac{1}{\sqrt{5}}h) + f(c+h)) 
\end{equation}
and furthermore a Kronrod formula can be used to estimate the precision of the integral\cite{patterson} 
\begin{eqnarray}
	Q' &=& \frac{h}{1470}(77f(c-h) + 432f(c-\sqrt{\frac{2}{3}}h) \nonumber\\
	&&+ 625f(c-\frac{1}{\sqrt{5}}h)+ 672f(c) \nonumber\\
	&&+ 625f(c+\frac{1}{\sqrt{5}}h) + 432f(c+\sqrt{\frac{2}{3}}h)\nonumber \\
	&&+ 77f(c+h)) 
\end{eqnarray}

The difference between $Q$ and $Q'$ can be taken as the precision of the integral.

The adaptive procedure to get the integral of the Green function in a region $[c-h,c+h]$  is
\begin{enumerate}
	\item calculate $Q$ and $Q'$, then compare the difference $\Delta$ with the tolerance $\delta$.\\
	\item if $\Delta$ is smaller than $\delta$, the integral is converged and $Q$ is used as integral result. If not, divide the region $[c-h,c+h]$ into three subregions $[c-h,c-\frac{1}{\sqrt{5}}h], [c-\frac{1}{\sqrt{5}}h, c+\frac{1}{\sqrt{5}}h], [c+\frac{1}{\sqrt{5}}h, c+h]$ and redo the step 1 for each subregion until the integral is converged in the whole region.\\
\end{enumerate}

For the lesser Green function inside the bias window, we use the simple
composite trapezoidal rule to obtain the integral. However, numerical
errors can easily occur close to the real axis where the
Green function has singularities. For this reason we apply the double contour method introduced in Ref.~\onlinecite{BrandbyNEGF}: First, the integral of the retarded Green
function is calculated along the path CD (Fig.~\ref{fig:contour})
above the bias window, which is the spectrum for all the electronic states 
in the bias window $S$, then both electron
spectrum $G^<$ and hole spectrum $G^>$ are integrated along the path
EF, and we have $S = D_e + D_h$ according to the definition of the Green
function, where electron density $D_e$ and hole density $D_h$ are
obtained from the integral of $G^<$ and $G^>$ respectively. The
numeric error, $\Delta S = S - D_e - D_h$ is normally a non-zero quantity due
to the integral insufficiency. As a correction, the error is
distributed to $D_e$ and $D_h$ by the matrix element weight.

\subsection{Sparse matrix handling} 
Because the matrix inversion cost scales as $N^3$,
where $N$ is the dimension of the matrix, the matrix inversion turns
out to be the main computational cost for large systems. Hence a sparse
matrix method is implemented to obtain the Green function.

We define a quenching layer as a slab whose left side has no overlap
with the right side due to the finite cutoff in the range of the
atomic orbitals. Hence an overlap or Hamiltonian matrix can be split
into several blocks, with each block representing the onsite values of
a quenching layer or the coupling between two adjacent quenching
layers. Note that quenching layers here are different from principal
layers used in the transport framework, where the latter is supposed
to be repeatable as well.

\begin{figure}
	[ht] \centering 
	\includegraphics[width=0.5
	\textwidth]{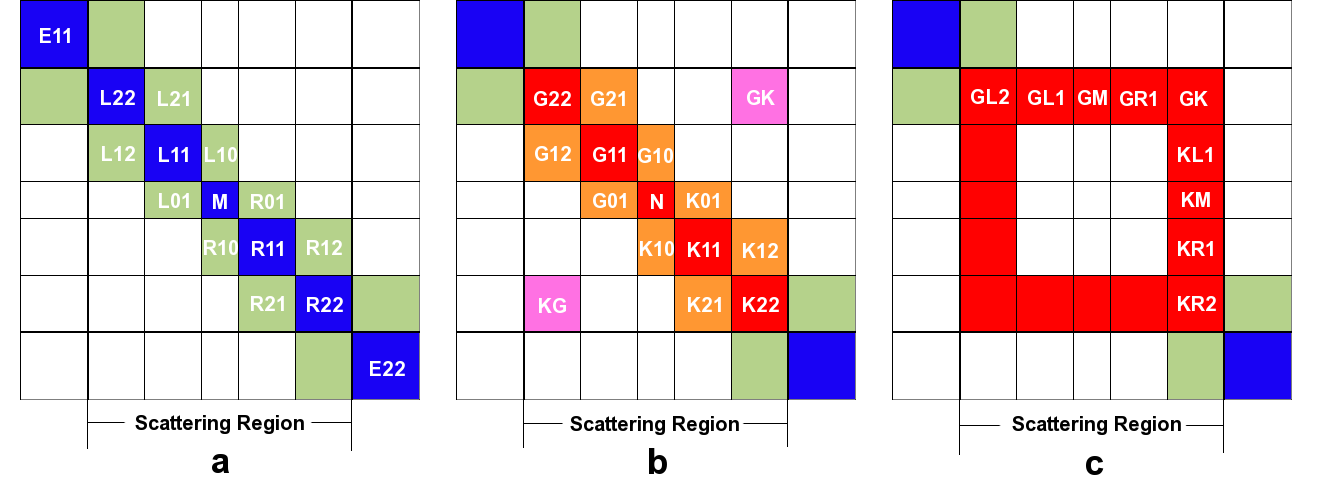}
	\caption{Schematic of the matrix blocks. A, the Hamiltonian or overlap matrix, the blue and green blocks represent the on-site and coupling sub-matrices for the different quenching layers respectively; B, the Green function matrix when evaluating the density, $DOS$ or transmission, the red and orange blocks represent the sub-matrices needed to calculate the density matrix or $DOS$, and pink blocks are for the transmission coefficient; C, The red blocks in the retarded Green function matrix are necessary when calculating the Keldysh Green function.}
\label{fig:matrix} 
\end{figure}

Physical quantities like density or transmission is often determined
by fairly few blocks of a matrix. To see this consider the simple
example of a two-probe system. In this case, the scattering region is
divided into 5 quenching layers. Fig.~\ref{fig:matrix} shows the
sparse matrix structure of the overlap or the Hamiltonian matrix. The
blocks outside the scattering region are from electrode calculations
and always fixed.

First we discuss how to obtain the real-space pseudo-density which can
be obtained by a projection of the pseudo-density matrix as in
Eq.~\ref{eq:density}. We see that if the states $\Phi_\nu
(\mathbf{r})$ and $\Phi_\mu(\mathbf{r})$ have no overlap, the
contribution from the pseudo-density matrix is zero, i.e., the white
blocks in Fig.~\ref{fig:matrix}a do not affect the density. So when we
calculate the density matrix from the integral of the Green' function,
only the blue and green blocks in the Green function matrix
(Fig.~\ref{fig:matrix}a) are necessary. There are really two different
parts, because two types of Green functions are involved when
calculating the density matrix: the equilibrium part and the
non-equilibrium part. We need the blue and green blocks of the
retarded Green function for the former and of the Keldysh Green
function for the latter. Through Eq.~\ref{eq:keldysh-green-function}
and the finite extent of the self-energy matrix, which is only
non-zero in the principle layers close to the electrodes, we see that
the blue and green blocks in Fig.~\ref{fig:matrix} in the Keldysh
Green function matrix can be obtained from only the red blocks of the
retarded Green function (Fig.~\ref{fig:matrix}c). So when we do the
matrix inversion to calculate the retarded Green function by
Eq.~\ref{eq:retarded-green-function}, the red blocks in
Fig.~\ref{fig:matrix}a are necessary for energy points on the path EF
in Fig.~\ref{fig:contour}, and the blue and green blocks in Fig.~\ref{fig:matrix}c
are necessary for energy points on the other path segments in
Fig.~\ref{fig:contour}.

We can also see that the red and orange blocks in Fig.~\ref{fig:matrix}b are needed to calculate the density of states ($DOS$) by $DOS(\varepsilon)=-\frac{1}{\pi}Im(\text{Tr}(G(\varepsilon^+)S)$ and the pink blocks in Fig.~\ref{fig:matrix}b are needed to calculate the transmission function $T(\varepsilon)=\text{Tr}[\Gamma_L(\varepsilon)G(\varepsilon^+)\Gamma_R(\varepsilon)G(\varepsilon^+)^{\dagger}]$.

The formulas below provide a quick solution for the necessary blocks. Here we just consider this particular matrix(shown in Fig.~\ref{fig:matrix})  as an example to show how the method works. A general formalism, which works for arbitrary number of electrodes and arbitrary number of principal layers in each electrode, is introduced in Ref.~\onlinecite{SparseMatrixQian}.

First, the central block $N$ in Fig.~\ref{fig:matrix}b of the retarded Green function can be solved through the equations 
\begin{eqnarray}
	Q^L_2 &=&L_{22}^{-1}; Q^L_1 =(L_{11}-L_{12}Q_2^LL_{21})^{-1}\\
	Q^R_2 &=&R_{22}^{-1}; Q^R_1 =(R_{11}-R_{12}Q_2^RR_{21})^{-1}\\
	N &=&(M-\sum_{J=L,R} J_{12}Q^J_2J_{21})^{-1},
\end{eqnarray}
where $L_{ij}$ and $R_{ij}$ are the blocks shown in Fig.~\ref{fig:matrix}a representing the matrix $\varepsilon S-H_{CC}-\Sigma^r_L(\varepsilon) - \Sigma^r_R(\varepsilon)$. Then, for the remaining blocks of the retarded Green function matrix, we have to iterate the formulas
\begin{eqnarray}
	G^L_{i,i} &=& Q_{i}^L(I-L_{i,i-1}G_{i-1,i}^L) \nonumber\\ 
	G^L_{i,i} &=& (I-G_{i,i-1}L_{i-1,i}^L)Q_{i}^L \nonumber\\
	G^L_{i,j} &=& -Q_{i}L_{i,i-1}G^L_{i-1,j} \nonumber \\
	G^L_{j,i} &=& -G^L_{j,i-1}L_{i-1,i}Q^L_{i} (j < i) ,\label{eq:block}
\end{eqnarray}
where $G^L_{i,j}$ is the block from the central part to electrode L and $G^L_{0,0}$ is $N$ in Eq.~\ref{eq:block}, the blocks from the central part to electrode R have a similar solution. For all the required blocks, a quick solution can be obtained using a combination of the recursive formulas Eq.~\ref{eq:block}. If we denote the number of quenching layers by $n$, the computational cost is roughly given by $n$ times the cost of an inversion operation plus 4$n$ times the cost of matrix multiplication\cite{SparseMatrixQian}.

\subsection{Fixed boundary conditions} 
The electronic potential of the metal electrodes will usually be very efficiently screened so that after only a few atomic layers into the electrodes we can assume that the potential is equal to the equilibrium potential plus/minus a possible constant bias potential. We shall apply open boundary conditions (in contrast to, say, periodic ones) where the bias is applied by fixing the potential values at the boundaries before solving the Poisson equation\cite{JapanNEGF}. This procedure also allows for a net charge in the scattering region in which case the perturbation of the electron potential into the electrodes will of course be somewhat more long ranged.

The Poisson equation $\nabla^2\tilde{v}_{coul}=-4\pi\tilde{\rho}$ is solved in reciprocal space in the $x$ and $y$ directions while it is solved in real space in the $z$-coordinate, i.e., in the the transport direction. Mathematically we have
\begin{equation}
	(\frac{d^2}{dz^2}-\vec{G}^2)\tilde{v}_{coul}(z, \vec{G})=-4\pi\tilde{\rho}(z, \vec{G}),
	\label{eq:fourier}
\end{equation}
where $\vec{G}$ is the vectors of the 2-d real grids used for the Fourier transformation.

Eq.~\ref{eq:fourier} is solved by the sparse matrix linear equation subroutine provided by the Lapack package. An advantage of this Poisson equation solution is the good parallelization behavior. The 2-d Fourier transformations are independent for the different real-space slices, and the linear equations Eq.~\ref{eq:fourier} can be solved independently for different $\vec{G}$-vectors.

\section{Results} 
\subsection{Quantum capacitor} 
\begin{figure}
	[ht] \centering 
	\includegraphics[width=0.5
	\textwidth]{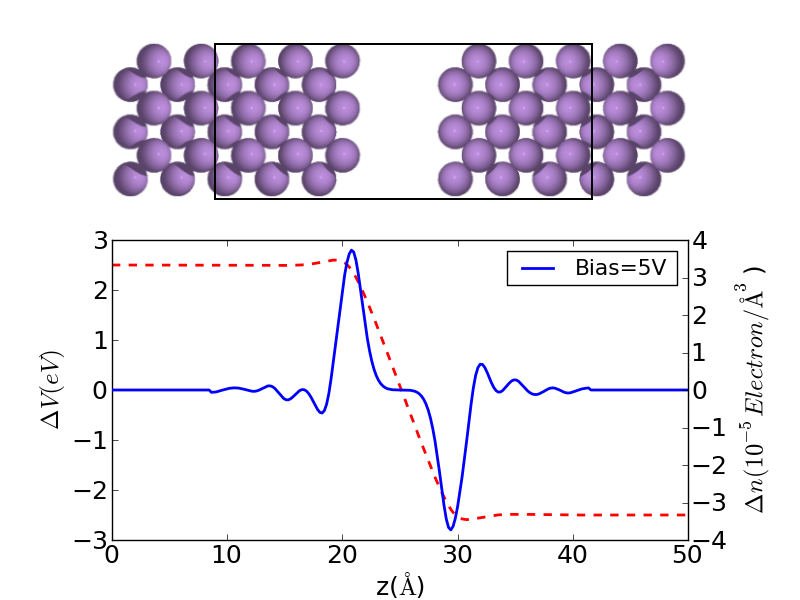} \caption{Upper panel: Capacitor model with the electrodes made of bcc sodium with the voltage drop along the (100) direction. The vacuum distance between the two electrodes is $8~{\text{\AA}}$, and the size of the unit cell in the transverse directions is $12.7{\text{\AA}}\times12.7{\text{\AA}}$. The rectangle represents the scattering region. Lower panel: The non-equilibrium part of the electron density (solid blue) and the induced effective potential (dashed red) under a bias voltage of 5V, with the zero bias values as the reference. The calculated values are averaged over the transverse plane.}
	\label{fig:capacitor} 
\end{figure}

We consider a simple
capacitor system consisting of two semi-infinite Na electrodes
separated by a vacuum gap, see upper part of
Fig.~\ref{fig:capacitor}. When a bias voltage is applied to the system, electrons accumulate/deplete on the
Na(100) surfaces. According to the classic parallel plate capacitor model, the surface
charge should be 
\begin{equation}
Q_{cl}=\varepsilon_0 VA/d,
\end{equation}
where $\varepsilon_0$, $A$, $d$ are the vacuum permittivity, area
cross section of the unit cell, and the gap distance, respectively. We
integrate the induced charge density in real space and obtain the net
charge accumulation $Q=0.45e$, which is close to the result of classic
theory $Q_{cl}=0.55e$. One difference is that in the quantum theory,
the charge decays from the surface into the bulk in an
oscillatory fashion (Friedel oscillations), instead of being localized
exactly on the surface as assumed in the classic model. We also note
that at a distance of about 4 layers from the surface the values for
both the potential and the charge are very close to their bulk values.
Hence this calculation confirms the screening approximation, namely
that a few layers away from the surface or scattering region the
potential has reached its bulk value. Finally, we note that the relatively high
bias voltage of $\approx 5V$ is possible in the present case where no current
is flowing. On the other hand, the non-equilibrium states determining the current flow become
increasingly difficult to calculate accurately for larger bias values due to the insufficient integral
of the Keldysh Green function in the bias window. As a consequence electron transport 
calculations are typically possible/reliable up to bias
voltages of around $2-3$V, depending on the transparency of the junction.

\subsection{Non-equilibrium forces} 
The calculation of non-equilibrium forces is in principle a delicate
problem involving non-conservative components\cite{Non_conservative_force,hedegaard,todorov}. For
highly conducting molecular bridges an instability may occur which
involves the Berry phase of the wave function. The
description of such phenomena is beyond the scope of the usual
NEGF+DFT framework, but in simpler cases, in particular in cases with no or little current flow,  
the force expressions
Eqs.~\ref{eq:force}-\ref{eq:force2} still apply.

\begin{figure}
	[ht] \centering 
	\includegraphics[width=.5
	\textwidth]{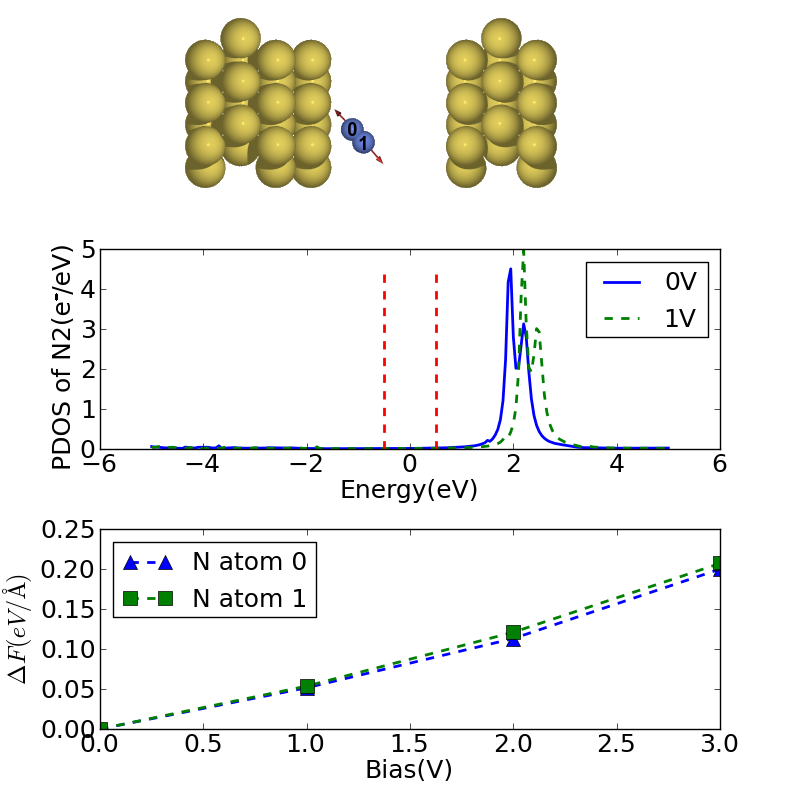} \caption{The bias voltage effect on a nitrogen molecule between two gold electrodes. Upper panel: the atomic structure of the system. The arrows represent the directions of the atomic forces generated by the bias voltage. Middle panel: $PDOS$(partial density of states) of the nitrogen molecule at 0V and 1V bias voltage. The Fermi level is located at zero and the red dashed lines show the location of the bias window in the 1V case. Lower panel: the magnitude of the non-equilibrium atomic forces as a function of bias voltage.}
	\label{fig:N2Au}
\end{figure}

As an example, we show here a non-equilibrium force calculation for
a Au/N$_2$/Au junction, where we can see the tendency towards molecular
dissociation under a bias voltage. The structure (see upper part of
Fig.~\ref{fig:N2Au}) is relaxed under zero bias until the maximum force is below 
0.01eV/${\text{\AA}}$. When a positive bias is applied,
electrons are redistributed over the molecule due to the electric field. Consequently, the two nitrogen
atoms start to repel each other due to increased Coulomb repulsion which weakens the bond. The actual
quantity of charge transfer to the molecule, which is about 0.01e for
1V bias voltage on this system, shifts up the molecular energy
spectrum, i.e. the energy levels follow the chemical potential of the left electrode (see Fig.~\ref{fig:N2Au} middle panel). The force is mainly occurring only between the
two nitrogen atoms, while there is no force induced between the electrode atoms and
the N$_2$ molecule. Equivalently, a negative bias voltage shifts down the levels and pull electrons out of the N$_2$ molecule, and leads to an attractive force between the two nitrogen atoms.

\subsection{Electrostatic gate control} 
One way of controlling the current flow through a nano-scale conductor is by electrostatic gating. This has been demonstrated experimentally for graphene, where a metal-insulator transition was induced by gating\cite{Gate_BilayerGNR, Gate_TrilayerGNR}, and for single-molecule junctions where the individual electronic levels were moved in energy relative to the Fermi level of the source/drain electrodes\cite{kubatkin,Gate_Mark_Reed}. At the single molecule scale, the gate-molecule coupling is to a large extent determined by the device
geometry with the screening effect playing an important
role\cite{Datta_gate}. For numerical simulations, the typical method
of applying a gate is to add an external potential $v_g(\bf r)$ to the effective potential of
the system
\begin{equation}
	\tilde{v}(\mathbf r) = \tilde{v}^0(\mathbf r) + v_g(\mathbf r). 
\end{equation}

\begin{figure}
	[ht] \centering 
	\includegraphics[width=0.5
	\textwidth]{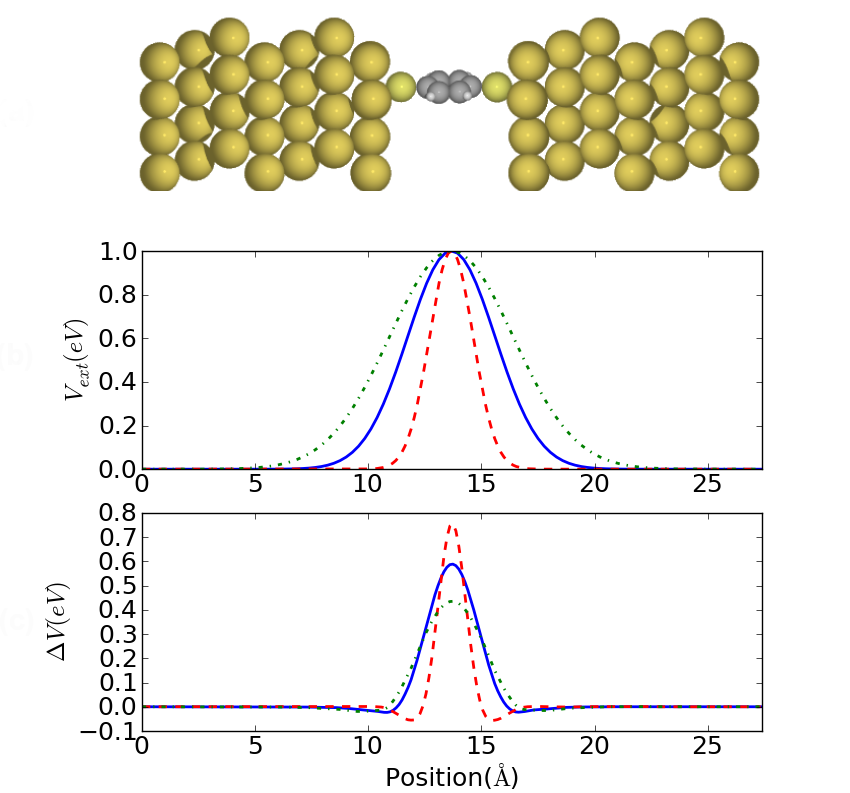} \caption{Effect of electrostatic gating of a benzene-1,4-dithiolate molecule between two gold fcc(111) electrodes. Upper panel: the atomic structure of the system. Middle panel: the applied external potential. Lower panel: the effective additional potential after self-consistency.\label{fig:gate}} 
\end{figure}

We now consider the prototypical Au-BDT-Au junction (see upper panel of Fig.~\ref{fig:gate}) subject to three different gate potentials, $v_g(z)$ (see middle panel of Fig.~\ref{fig:gate}). We note that the structure of the Au/BDT junction is presently being debated\cite{hakkinen,jadzinsky,ulstrup}. However, for our purpose the simple model makes the sense. 

In the lower part of Fig.~\ref{fig:gate} we plot the resulting effective gate potential $\Delta v(z)=\tilde{v}_{sc}(z)- \tilde{v}^0_{sc}(z)$ where the subscript $sc$ denotes self-consistency, and the superscript $0$ denotes zero gate voltage. Due to the screening in the metal, the effective gate potentials only affect the molecule region, and the narrower gate potential is seen to be less influenced by the self-consistency because it does not induce considerable charge transfer at the metal surfaces - a charge transfer that otherwise tends to reduce the gate effect on the molecule. We note that the gate efficiency factor, $\alpha=\Delta v_(z)/v_g(z)$, for these three potentials are about 0.8, 0.6, and 0.4 in the molecular region with the larger efficiency obtained for the more localized gate potential. The value of 0.4 obtained for the most delocalized potential is fairly close to an experimental study\cite{Gate_Mark_Reed} of the Au-BDT-Au system where an efficiency factor of 0.25 was reported. 

\begin{figure}
	[ht] \centering 
	\includegraphics[width=0.5
	\textwidth]{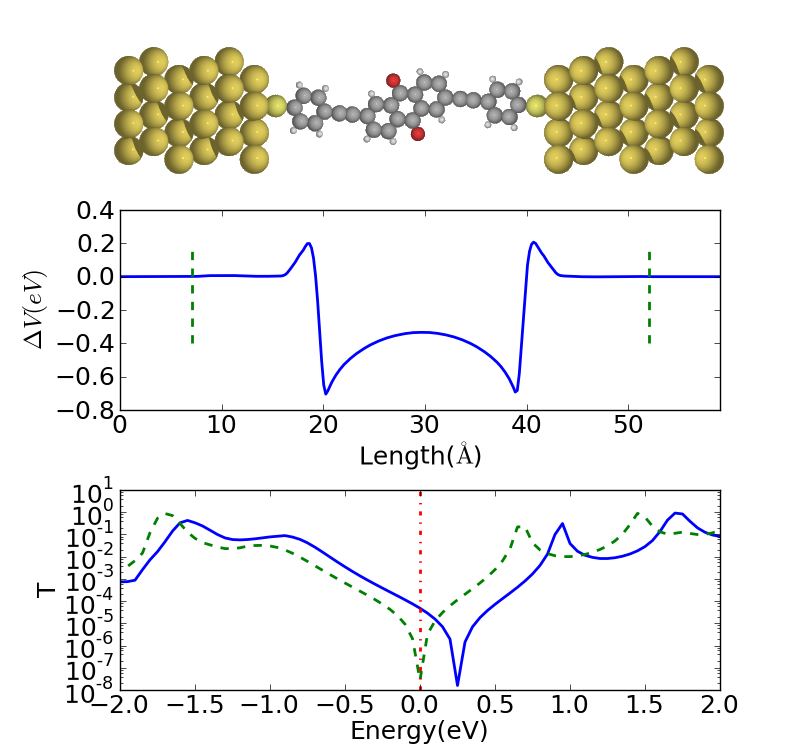} \caption{Gate-tuning the conductance of a molecular transistor.  Upper panel: the Au-anthraquinone-Au structure. Middle panel: the gate effect on the potential. Lower panel: the transmission coefficient at 0V and -2V gate voltages.\label{fig:anth}} 
\end{figure}

In the following we illustrate how the gate voltage can be used to tune the conductance of a molecular junction. It has recently been shown that the transport through the molecule anthraquinone is strongly suppressed due to destructive quantum interference occurring close to the Fermi level when the molecule is connected to metallic electrodes\cite{TroelsQI,troels2}. The quantum interference leads to a dip in the transmission function inside the energy gap between the highest occupied (HOMO) and lowest unoccupied (LUMO) molecular orbitals. Hence a large on-off ratio is
expected when shifting the molecular energy levels by an external gate
voltage. The upper panel of Fig.~\ref{fig:anth} shows the molecule
connected to two gold fcc(111) surfaces through Au-S bonds. The
effective potential with the gate voltage -2V is shown in the middle panel. 
We see that the potential of the central part of the
anthraquinone molecule is shifted less than the potential for the
outer parts of the molecule. This is due to the fact that different
parts of the molecule polarize differently as a consequence of the
detailed electronic structure. The HOMO is for example mainly
localized at the connecting wires. The lower panel of Fig.~\ref{fig:anth} shows the change of transmission coefficient when a gate voltage of -2V is applied. Due to the characteristic interference dip in the transmission a large on-off ratio of about a factor of 1000 is achieved. Note that the relatively poor gate efficiency of around 0.1 is due to Fermi level pining of the HOMO. 

\begin{figure}
	[ht] \centering 
	\includegraphics[width=0.5
	\textwidth]{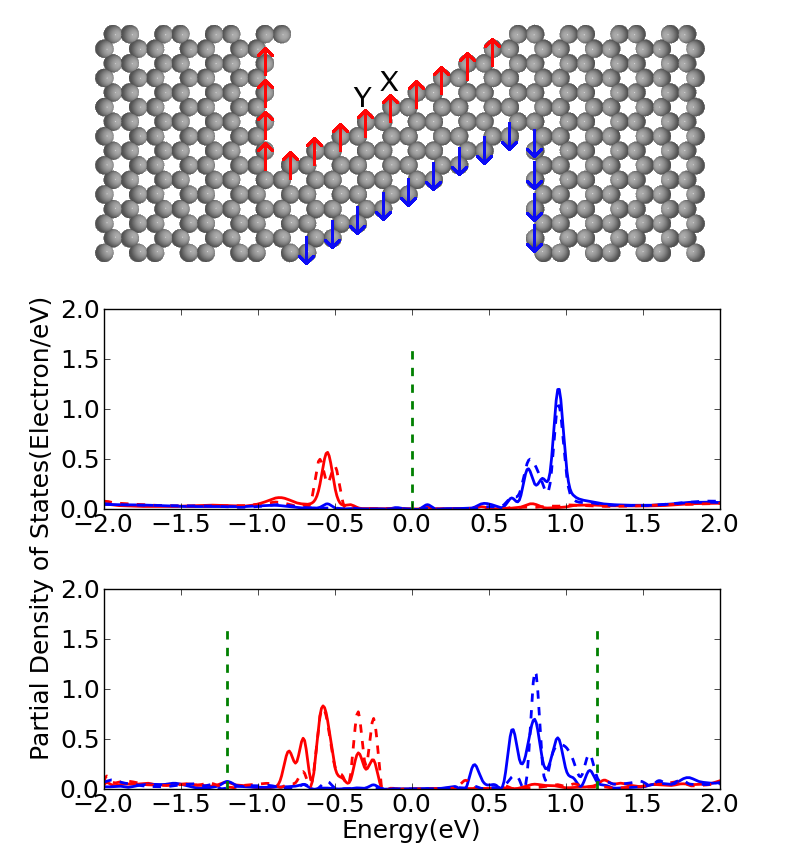} \caption{Spin transport in a zigzag graphene nanoribbon(ZGNR) bridge connecting to two semi-infinite graphene flakes.Upper panel: the atomic structure of the graphene/ZGNR/graphene system. Middle panel: the $PDOS$ of a C atom at the ribbon's zigzag edge under zero bias, the red and blue solid lines represent the spin-up and spin-down $PDOS$ of the C atom at the center of the zigzag edge(marked with X), the red and blue dashed lines represent the spin-up and spin-down $PDOS$ of the C atom next to the previous one at the zigzag edge(marked with Y), the green dashed line is the fermi level. Lower panel: the $PDOS$ of a C atom at the ribbon's zigzag edge under bias voltage $V=2.4V$.}
	\label{fig:spintransport}
\end{figure}

\subsection{Spin transport}
\begin{figure}
	[ht] \centering 
	\includegraphics[width=0.5
	\textwidth]{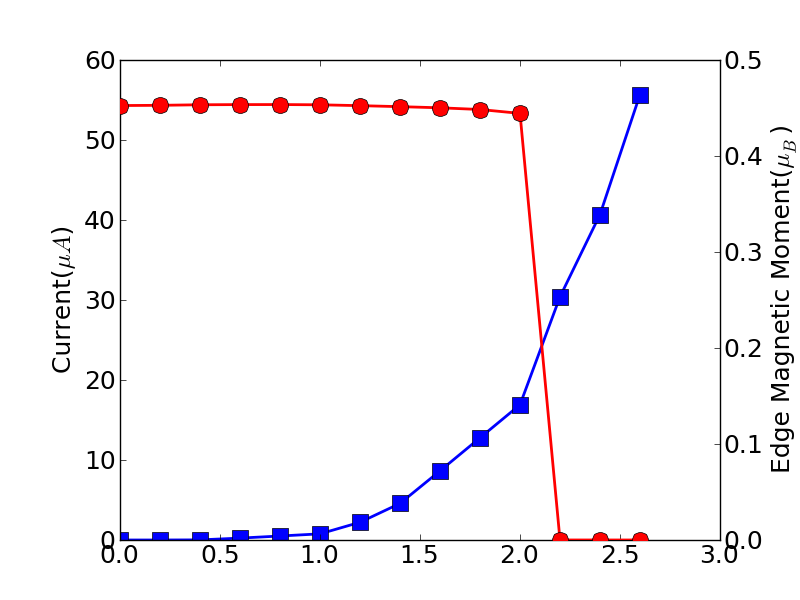} \caption{Calculated current (blue squares) and edge magnetic moment per C atom (red circles) as a function of bias voltage for the ZGNR structure shown in Fig. \ref{fig:spintransport}. }
\label{fig:IV_M}
\end{figure}

In this section, we investigate the nonequilibrium-driven magnetic transition 
in the spin transport in zigzag graphene nanoribbon(ZGNR)
which is proposed in Ref.\cite{Branislav} based on tight-binding
calculations.  The ZGNR's edge is spin-polarized and it has an
anti-ferromagnetic spin configuration if its number of atomic layers
is even\cite{spin_polarized_edge_state}. A gap about 1eV is opened between
the different spin states and makes the ZGNR a semiconductor. It was
noticed by Denis \emph{et al.} that the ZGNR's magnetic ordering is killed 
when the external bias voltage exceeds the size of the gap\cite{Branislav}. 
Here we reproduce this result for the graphene/ZGNR/graphene system shown 
in the upper panel of Fig.~\ref{fig:spintransport}, where a ZGNR(nn=8) is sandwiched
between two semi-infinite graphene flakes. Under zero bias the $PDOS$ of
the central C atom along the ZGNR edge shows two peaks above and below the Fermi
level, corresponding to the different spin directions (middle panel in
Fig.~\ref{fig:spintransport}). The distance between the two peaks is
about $1.6eV$, and equals the band gap. When this bias voltage reaches
$1.0eV$, the current starts to increase, see Fig.~\ref{fig:IV_M}. At
bias voltages $2.0eV$ the edge magnetic moment disappears very
abruptly, and the current starts to increase even faster.

Interestingly, in the tight-binding calculations presented in
Ref.~\cite{Branislav}, both the magnetic moment and the current show a
very abrupt feature at the bias threshold, while in our calculation,
the current increases rather smoothly. This can be explained by the
non-equilibrium potential in the ab-initio calculation leads to a
rehybridization and broadening of the spectral peaks, see lower panel
of Fig.~\ref{fig:spintransport}. We also note that in our calculation 
the disappearance of the magnetic moment is due to the two Stoner peaks moving
into the bias window being half-occupied, different from the complete 
band collapse in the Ref.~\cite{Branislav},this is because our ZGNR is not
long enough. We can see from the lower panel of Fig.~\ref{fig:spintransport}
the gap shrinks more for the C atom further from the contact.

\subsection{Multi-terminal transport} 
The expression for the Green function of the scattering region Eq.~(\ref{eq:retarded-green-function}) can be straightforwardly extended to a multi-terminal situation
\begin{equation}
	\label{eq:retarded-green-function_multi_terminal} G^r(\varepsilon) = [\varepsilon S-H_{CC}-\sum_{J}\Sigma^r_J(\varepsilon)]^{-1} 
\end{equation}
where $J$ is the index of the terminals. In contrast to the situation
for a two-probe calculation, a zero boundary condition is applied for
the effective potential for a multi-terminal system, and buffer atoms
are used to represent semi-infinite leads. This approach to
multiterminal transport has been previously investigated in
Ref.~\onlinecite{zhangjiaxing}. It should be noted that the self-energy
of a lead has to be ``rotated'' by an orthogonal transformation when
the lead is not along either the $x$, $y$ or $z$ axes. 

\begin{figure}
	[ht] \centering 
	\includegraphics[width=0.5
	\textwidth]{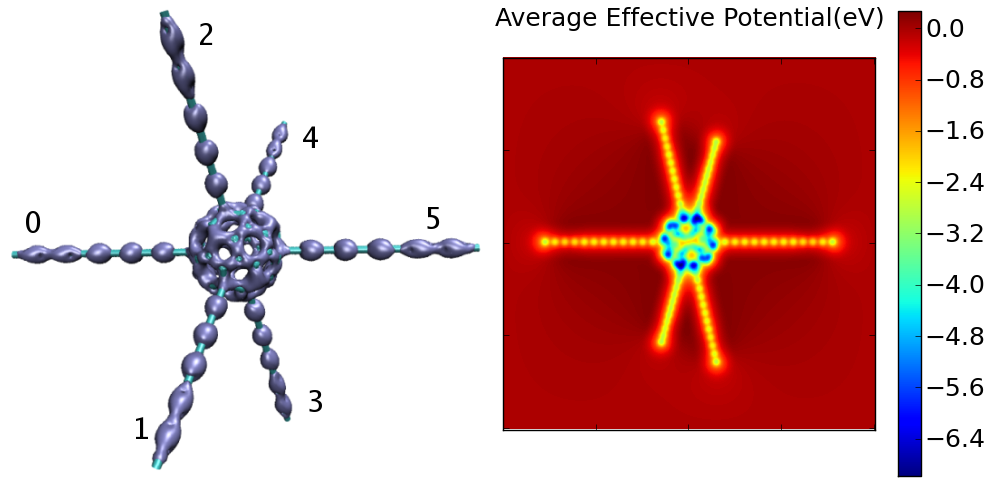} \caption{Left: The C$_{60}$-6-terminal structure and real space $DOS$ at the Fermi level. Right: The averaged effective potential projected onto a plane cutting through the C60 molecule.\label{fig:C60}} 
      \end{figure}

As an example, we consider a C$_{60}$ molecule connected to six linear
carbon atomic chains. Fig.~\ref{fig:C60}(left) shows the projected $DOS$
in real space evaluated at the Fermi level. The coverage suggests the
scattering states are itinerant in the whole system and the contact
between the carbon chain and the C$_{60}$ moelcule is transparent.  Fig.
\ref{fig:C60}(right) shows a 2D averaged potential in a plane cutting
through the C$_{60}$ molecule.

\begin{figure}
	[ht] \centering 
	\includegraphics[width=0.3
	\textwidth]{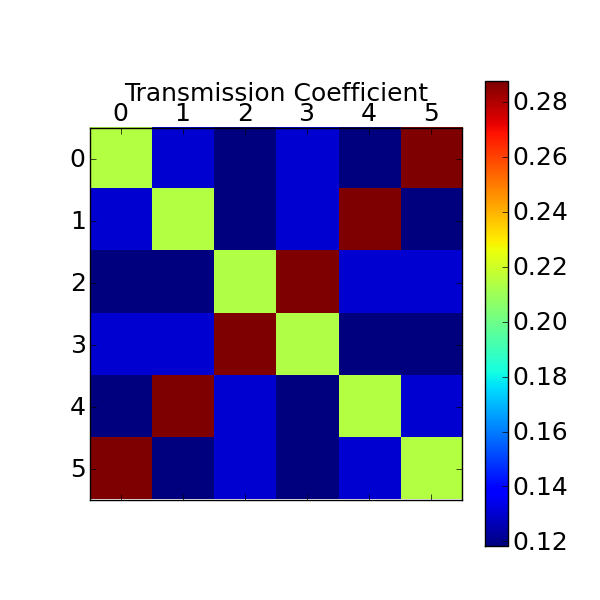} \caption{The transmission function evaluated at the Fermi energy between the six leads. The element $(i,j)$ refers to the transmission from lead $i$ to lead $j$; the diagonal element is the reflection coefficient.\label{fig:C60_trans}} 
\end{figure}

A matrix indicating the transmission at the Fermi level between the different leads is shown in the Fig.~\ref{fig:C60_trans}. The matrix index notation represents the lead number as shown in Fig. \ref{fig:C60}(left). In particular, the diagonal corresponds to back scattering, i.e. it gives the reflection probability. We can see that electrons are more easily transmitted between leads opposing each other, whereas the transmission decreases if the electron has to turn an angle during the scattering process.
This intuitive phenomenon can be explained by the quantum interference of the different partial waves. For the straightforward scattering, the quantum phases are the same for all the paths passing through the C$_{60}$ molecule, and the electron therefore attains the greatest transmission probability\cite{troels2}.

\section{Computational details} For completeness we list the key input parameters and CPU timings for the examples presented in this paper in the table below. 
\begin{table*}
	\begin{tabular}
		{|c|c|c|c|c|c|c|c} \hline System & K-sampling & Functional & Basis & Lead K-sampling & processor number & walltime(hour)\\
		\hline Capacitor & (4,4) & LDA & SZ & (4,4,15) & 4 & 0.5\\
		\hline Au/N2/Au & (2,2) & PBE & DZP(SZP) & (2,2,15)& 32 & 3 \\
		\hline Gate(BDT) & (2,2) & PBE &DZP(SZP) & (2,2,15) & 32 & 6 \\
		\hline Gate(anthraquinon) & (2,2) & PBE &DZP(SZP) & (2,2,15) & 32 & 8 \\
		\hline Graphene/ZGNR/Graphene & (2,1) & PBE & SZP & (2,2,7) & 32 & 6 \\
		\hline C60 & (1,1) & LDA & SZ & (1,1,15) & 32 & 2\\
		\hline 
	\end{tabular}
	\caption{Key parameters and CPU timings for the examples considered in this article.}
	\label{tab:systems}
\end{table*}

\section{Summary} We have described the implementation of the NFGF+DFT
transport method in the GPAW code and illustrated its performance
through application to a number of different molecular junctions.  The
electronic structure is described within the PAW methodology which
provides all-electron accuracy at a computational cost corresponding to
pseudopotential calculations. The Green functions are represented in a
basis set consisting of atomic-like orbitals while the Poisson
equation with appropriate boundary conditions is solved in real space.
The code is parallelized over k-points and real space domains and
sparse matrix techniques are applied for maximal efficiency. The
flexibility of the method was illustrated through examples
demonstrating electron transport under finite bias voltage, effect of
electrostatic gating, spin transport, nonequilibrium forces, and
multi-terminal leads.

\section{Acknowledgment} The authors thank Jens J{\o}rgen Mortensen and Ask Hjorth Larsen for helpful discussions. The authors acknowledge support from the Danish Center for Scientific Computing through grant HDW-1103-06. The Center for Atomic-scale Materials Design is sponsored by the Lundbeck Foundation.


\begin{thebibliography}{52}
\expandafter\ifx\csname natexlab\endcsname\relax\def\natexlab#1{#1}\fi
\expandafter\ifx\csname bibnamefont\endcsname\relax
  \def\bibnamefont#1{#1}\fi
\expandafter\ifx\csname bibfnamefont\endcsname\relax
  \def\bibfnamefont#1{#1}\fi
\expandafter\ifx\csname citenamefont\endcsname\relax
  \def\citenamefont#1{#1}\fi
\providecommand{\bibinfo}[2]{#2}

\bibitem[{\citenamefont{Smit et~al.}(2002)\citenamefont{Smit, Noat, Untiedt,
  Lang, van Hemert, and van Ruitenbeek}}]{smit02}
\bibinfo{author}{\bibfnamefont{R.~H.~M.} \bibnamefont{Smit}},
  \bibinfo{author}{\bibfnamefont{Y.}~\bibnamefont{Noat}},
  \bibinfo{author}{\bibfnamefont{C.}~\bibnamefont{Untiedt}},
  \bibinfo{author}{\bibfnamefont{N.~D.} \bibnamefont{Lang}},
  \bibinfo{author}{\bibfnamefont{M.~C.} \bibnamefont{van Hemert}},
  \bibnamefont{and} \bibinfo{author}{\bibfnamefont{J.~M.} \bibnamefont{van
  Ruitenbeek}}, \bibinfo{journal}{Nature}, \textbf{\bibinfo{volume}{419}}, \bibinfo{pages}{906}
  (\bibinfo{year}{2002}).

\bibitem[{\citenamefont{Djukic et~al.}(2005)\citenamefont{Djukic, Thygesen,
  Untiedt, Smit, Jacobsen, and van Ruitenbeek}}]{djukic}
\bibinfo{author}{\bibfnamefont{D.}~\bibnamefont{Djukic}},
  \bibinfo{author}{\bibfnamefont{K.~S.} \bibnamefont{Thygesen}},
  \bibinfo{author}{\bibfnamefont{C.}~\bibnamefont{Untiedt}},
  \bibinfo{author}{\bibfnamefont{R.~H.~M.} \bibnamefont{Smit}},
  \bibinfo{author}{\bibfnamefont{K.~W.} \bibnamefont{Jacobsen}},
  \bibnamefont{and} \bibinfo{author}{\bibfnamefont{J.~M.} \bibnamefont{van
  Ruitenbeek}}, \bibinfo{journal}{Phys. Rev. B} \textbf{\bibinfo{volume}{71}},
  \bibinfo{pages}{161402} (\bibinfo{year}{2005}).

\bibitem[{\citenamefont{Venkataraman et~al.}(2006)\citenamefont{Venkataraman,
  Klare, Tam, Nuckolls, Hybertsen, and Steigerwald}}]{venkataraman}
\bibinfo{author}{\bibfnamefont{L.}~\bibnamefont{Venkataraman}},
  \bibinfo{author}{\bibfnamefont{J.~E.} \bibnamefont{Klare}},
  \bibinfo{author}{\bibfnamefont{I.~W.} \bibnamefont{Tam}},
  \bibinfo{author}{\bibfnamefont{C.}~\bibnamefont{Nuckolls}},
  \bibinfo{author}{\bibfnamefont{M.~S.} \bibnamefont{Hybertsen}},
  \bibnamefont{and} \bibinfo{author}{\bibfnamefont{M.~L.}
  \bibnamefont{Steigerwald}}, \bibinfo{journal}{Nano Lett.}
  \textbf{\bibinfo{volume}{6}}, \bibinfo{pages}{458} (\bibinfo{year}{2006}).

\bibitem[{\citenamefont{Quek et~al.}(2007)\citenamefont{Quek, Venkataraman,
  Choi, Louie, Hybertsen, and Neaton}}]{quek}
\bibinfo{author}{\bibfnamefont{S.~Y.} \bibnamefont{Quek}},
  \bibinfo{author}{\bibfnamefont{L.}~\bibnamefont{Venkataraman}},
  \bibinfo{author}{\bibfnamefont{H.~J.} \bibnamefont{Choi}},
  \bibinfo{author}{\bibfnamefont{S.~G.} \bibnamefont{Louie}},
  \bibinfo{author}{\bibfnamefont{M.~S.} \bibnamefont{Hybertsen}},
  \bibnamefont{and} \bibinfo{author}{\bibfnamefont{J.~B.}
  \bibnamefont{Neaton}}, \bibinfo{journal}{Nano Lett.}
  \textbf{\bibinfo{volume}{7}}, \bibinfo{pages}{3477} (\bibinfo{year}{2007}).

\bibitem[{\citenamefont{Li et~al.}(2007)\citenamefont{Li, Pobelov, Wandlowski,
  Bagrets, Arnold, and Evers}}]{evers}
\bibinfo{author}{\bibfnamefont{C.}~\bibnamefont{Li}},
  \bibinfo{author}{\bibfnamefont{I.}~\bibnamefont{Pobelov}},
  \bibinfo{author}{\bibfnamefont{T.}~\bibnamefont{Wandlowski}},
  \bibinfo{author}{\bibfnamefont{A.}~\bibnamefont{Bagrets}},
  \bibinfo{author}{\bibfnamefont{A.}~\bibnamefont{Arnold}}, \bibnamefont{and}
  \bibinfo{author}{\bibfnamefont{F.}~\bibnamefont{Evers}}, \bibinfo{journal}{J.
  Am. Chem. Soc.} \textbf{\bibinfo{volume}{130}}, \bibinfo{pages}{318}
  (\bibinfo{year}{2007}). 

\bibitem[{\citenamefont{\ifmmode~\mbox{\c{S}}\else \c{S}\fi{}ahin and
  Senger}(2008)}]{sahin}
\bibinfo{author}{\bibfnamefont{H.}~\bibnamefont{\ifmmode~\mbox{\c{S}}\else
  \c{S}\fi{}ahin}} \bibnamefont{and} \bibinfo{author}{\bibfnamefont{R.~T.}
  \bibnamefont{Senger}}, \bibinfo{journal}{Phys. Rev. B}
  \textbf{\bibinfo{volume}{78}}, \bibinfo{pages}{205423}
  (\bibinfo{year}{2008}).

\bibitem[{\citenamefont{Novaes et~al.}(2010)\citenamefont{Novaes, Rurali, and
  Ordej\'{o}n}}]{novaes}
\bibinfo{author}{\bibfnamefont{F.~D.} \bibnamefont{Novaes}},
  \bibinfo{author}{\bibfnamefont{R.}~\bibnamefont{Rurali}}, \bibnamefont{and}
  \bibinfo{author}{\bibfnamefont{P.}~\bibnamefont{Ordej\'{o}n}},
  \bibinfo{journal}{ACS Nano} \textbf{\bibinfo{volume}{4}},
  \bibinfo{pages}{7596} (\bibinfo{year}{2010}).

\bibitem[{\citenamefont{Markussen
  et~al.}(2010{\natexlab{a}})\citenamefont{Markussen, Rurali, Cartoix\`a,
  Jauho, and Brandbyge}}]{markussen}
\bibinfo{author}{\bibfnamefont{T.}~\bibnamefont{Markussen}},
  \bibinfo{author}{\bibfnamefont{R.}~\bibnamefont{Rurali}},
  \bibinfo{author}{\bibfnamefont{X.}~\bibnamefont{Cartoix\`a}},
  \bibinfo{author}{\bibfnamefont{A.-P.} \bibnamefont{Jauho}}, \bibnamefont{and}
  \bibinfo{author}{\bibfnamefont{M.}~\bibnamefont{Brandbyge}},
  \bibinfo{journal}{Phys. Rev. B} \textbf{\bibinfo{volume}{81}},
  \bibinfo{pages}{125307} (\bibinfo{year}{2010}{\natexlab{a}}).

\bibitem[{\citenamefont{Lee et~al.}(2010)\citenamefont{Lee, Kakushima,
  Shiraishi, Natori, and Iwai}}]{lee}
\bibinfo{author}{\bibfnamefont{Y.}~\bibnamefont{Lee}},
  \bibinfo{author}{\bibfnamefont{K.}~\bibnamefont{Kakushima}},
  \bibinfo{author}{\bibfnamefont{K.}~\bibnamefont{Shiraishi}},
  \bibinfo{author}{\bibfnamefont{K.}~\bibnamefont{Natori}}, \bibnamefont{and}
  \bibinfo{author}{\bibfnamefont{H.}~\bibnamefont{Iwai}},
  \bibinfo{journal}{Journal of Applied Physics} \textbf{\bibinfo{volume}{107}},
  \bibinfo{pages}{113705} (\bibinfo{year}{2010}).

\bibitem[{\citenamefont{Ono and Hirose}(2005)}]{hirose}
\bibinfo{author}{\bibfnamefont{T.}~\bibnamefont{Ono}} \bibnamefont{and}
  \bibinfo{author}{\bibfnamefont{K.}~\bibnamefont{Hirose}},
  \bibinfo{journal}{Phys. Rev. Lett.} \textbf{\bibinfo{volume}{94}},
  \bibinfo{pages}{206806} (\bibinfo{year}{2005}).

\bibitem[{\citenamefont{Rungger et~al.}(2009)\citenamefont{Rungger, Mryasov,
  and Sanvito}}]{rungger}
\bibinfo{author}{\bibfnamefont{I.}~\bibnamefont{Rungger}},
  \bibinfo{author}{\bibfnamefont{O.}~\bibnamefont{Mryasov}}, \bibnamefont{and}
  \bibinfo{author}{\bibfnamefont{S.}~\bibnamefont{Sanvito}},
  \bibinfo{journal}{Phys. Rev. B} \textbf{\bibinfo{volume}{79}},
  \bibinfo{pages}{094414} (\bibinfo{year}{2009}).

\bibitem[{\citenamefont{Chen et~al.}(2011)\citenamefont{Chen, Hummelsh{\o}j,
  Thygesen, Myrdal, N{\o}rskov, and Vegge}}]{chen}
\bibinfo{author}{\bibfnamefont{J.}~\bibnamefont{Chen}},
  \bibinfo{author}{\bibfnamefont{J.~S.} \bibnamefont{Hummelsh{\o}j}},
  \bibinfo{author}{\bibfnamefont{K.~S.} \bibnamefont{Thygesen}},
  \bibinfo{author}{\bibfnamefont{J.~S.} \bibnamefont{Myrdal}},
  \bibinfo{author}{\bibfnamefont{J.~K.} \bibnamefont{N{\o}rskov}},
  \bibnamefont{and} \bibinfo{author}{\bibfnamefont{T.}~\bibnamefont{Vegge}},
  \bibinfo{journal}{Catalysis Today} \textbf{\bibinfo{volume}{165}},
  \bibinfo{pages}{2 } (\bibinfo{year}{2011}).

\bibitem[{\citenamefont{Dell'Angela et~al.}(2010)\citenamefont{Dell'Angela,
  Kladnik, Cossaro, Verdini, Kamenetska, Tamblyn, Quek, Neaton, Cvetko,
  Morgante et~al.}}]{ref3}
\bibinfo{author}{\bibfnamefont{M.}~\bibnamefont{Dell'Angela}},
  \bibinfo{author}{\bibfnamefont{G.}~\bibnamefont{Kladnik}},
  \bibinfo{author}{\bibfnamefont{A.}~\bibnamefont{Cossaro}},
  \bibinfo{author}{\bibfnamefont{A.}~\bibnamefont{Verdini}},
  \bibinfo{author}{\bibfnamefont{M.}~\bibnamefont{Kamenetska}},
  \bibinfo{author}{\bibfnamefont{I.}~\bibnamefont{Tamblyn}},
  \bibinfo{author}{\bibfnamefont{S.~Y.} \bibnamefont{Quek}},
  \bibinfo{author}{\bibfnamefont{J.~B.} \bibnamefont{Neaton}},
  \bibinfo{author}{\bibfnamefont{D.}~\bibnamefont{Cvetko}},
  \bibinfo{author}{\bibfnamefont{A.}~\bibnamefont{Morgante}},
  \bibnamefont{et~al.}, \bibinfo{journal}{Nano Lett.}
  \textbf{\bibinfo{volume}{10}}, \bibinfo{pages}{2470} (\bibinfo{year}{2010}).

\bibitem[{\citenamefont{Garcia-Lastra et~al.}(2009)\citenamefont{Garcia-Lastra,
  Rostgaard, Rubio, and Thygesen}}]{juanma}
\bibinfo{author}{\bibfnamefont{J.~M.} \bibnamefont{Garcia-Lastra}},
  \bibinfo{author}{\bibfnamefont{C.}~\bibnamefont{Rostgaard}},
  \bibinfo{author}{\bibfnamefont{A.}~\bibnamefont{Rubio}}, \bibnamefont{and}
  \bibinfo{author}{\bibfnamefont{K.~S.} \bibnamefont{Thygesen}},
  \bibinfo{journal}{Phys. Rev. B} \textbf{\bibinfo{volume}{80}},
  \bibinfo{pages}{245427} (\bibinfo{year}{2009}).

\bibitem[{\citenamefont{Toher and Sanvito}(2007)}]{toher}
\bibinfo{author}{\bibfnamefont{C.}~\bibnamefont{Toher}} \bibnamefont{and}
  \bibinfo{author}{\bibfnamefont{S.}~\bibnamefont{Sanvito}},
  \bibinfo{journal}{Phys. Rev. Lett.} \textbf{\bibinfo{volume}{99}},
  \bibinfo{pages}{056801} (\bibinfo{year}{2007}).

\bibitem[{\citenamefont{Strange et~al.}(2011)\citenamefont{Strange, Rostgaard,
  H\"akkinen, and Thygesen}}]{strange}
\bibinfo{author}{\bibfnamefont{M.}~\bibnamefont{Strange}},
  \bibinfo{author}{\bibfnamefont{C.}~\bibnamefont{Rostgaard}},
  \bibinfo{author}{\bibfnamefont{H.}~\bibnamefont{H\"akkinen}},
  \bibnamefont{and} \bibinfo{author}{\bibfnamefont{K.~S.}
  \bibnamefont{Thygesen}}, \bibinfo{journal}{Phys. Rev. B}
  \textbf{\bibinfo{volume}{83}}, \bibinfo{pages}{115108}
  (\bibinfo{year}{2011}).

\bibitem[{\citenamefont{Strange and Thygesen}(2011)}]{strange2}
\bibinfo{author}{\bibfnamefont{M.}~\bibnamefont{Strange}} \bibnamefont{and}
  \bibinfo{author}{\bibfnamefont{K.~S.} \bibnamefont{Thygesen}},
  \bibinfo{journal}{Beilstein Journal of Nanotechnology}
  \textbf{\bibinfo{volume}{2}}, \bibinfo{pages}{746} (\bibinfo{year}{2011}).

\bibitem[{\citenamefont{Rangel et~al.}(2011)\citenamefont{Rangel, Ferretti,
  Trevisanutto, Olevano, and Rignanese}}]{rangel}
\bibinfo{author}{\bibfnamefont{T.}~\bibnamefont{Rangel}},
  \bibinfo{author}{\bibfnamefont{A.}~\bibnamefont{Ferretti}},
  \bibinfo{author}{\bibfnamefont{P.~E.} \bibnamefont{Trevisanutto}},
  \bibinfo{author}{\bibfnamefont{V.}~\bibnamefont{Olevano}}, \bibnamefont{and}
  \bibinfo{author}{\bibfnamefont{G.-M.} \bibnamefont{Rignanese}},
  \bibinfo{journal}{Phys. Rev. B} \textbf{\bibinfo{volume}{84}},
  \bibinfo{pages}{045426} (\bibinfo{year}{2011}).

\bibitem[{\citenamefont{Hybertsen et~al.}(2008)\citenamefont{Hybertsen,
  Venkataraman, Klare, Whalley, Steigerwald, and Nuckolls}}]{hybertsen}
\bibinfo{author}{\bibfnamefont{M.~S.} \bibnamefont{Hybertsen}},
  \bibinfo{author}{\bibfnamefont{L.}~\bibnamefont{Venkataraman}},
  \bibinfo{author}{\bibfnamefont{J.~E.} \bibnamefont{Klare}},
  \bibinfo{author}{\bibfnamefont{A.~C.} \bibnamefont{Whalley}},
  \bibinfo{author}{\bibfnamefont{M.~L.} \bibnamefont{Steigerwald}},
  \bibnamefont{and} \bibinfo{author}{\bibfnamefont{C.}~\bibnamefont{Nuckolls}},
  \bibinfo{journal}{Journal of Physics: Condensed Matter}
  \textbf{\bibinfo{volume}{20}}, \bibinfo{pages}{374115}
  (\bibinfo{year}{2008}).

\bibitem[{\citenamefont{Stefanucci and Almbladh}(2004)}]{stefanucci_almbladh}
\bibinfo{author}{\bibfnamefont{G.}~\bibnamefont{Stefanucci}} \bibnamefont{and}
  \bibinfo{author}{\bibfnamefont{C.~O.} \bibnamefont{Almbladh}},
  \bibinfo{journal}{Europhys. Lett.},\textbf{\bibinfo{volume}{67}}, \bibinfo{pages}{14}(\bibinfo{year}{2004}).

\bibitem[{\citenamefont{Yam et~al.}(2011)\citenamefont{Yam, Zheng, Chen, Wang,
  Frauenheim, and Niehaus}}]{niehaus}
\bibinfo{author}{\bibfnamefont{C.~Y.}~\bibnamefont{Yam}},
  \bibinfo{author}{\bibfnamefont{X.}~\bibnamefont{Zheng}},
  \bibinfo{author}{\bibfnamefont{G.~H.}~\bibnamefont{Chen}},
  \bibinfo{author}{\bibfnamefont{Y.}~\bibnamefont{Wang}},
  \bibinfo{author}{\bibfnamefont{T.}~\bibnamefont{Frauenheim}},
  \bibnamefont{and} \bibinfo{author}{\bibfnamefont{T.~A.}
  \bibnamefont{Niehaus}}, \bibinfo{journal}{Phys. Rev. B}
  \textbf{\bibinfo{volume}{83}}, \bibinfo{pages}{245448}
  (\bibinfo{year}{2011}).

\bibitem[{\citenamefont{Dundas et~al.}(2009)\citenamefont{Dundas, McEniry, and
  N.}}]{todorov}
\bibinfo{author}{\bibfnamefont{D.}~\bibnamefont{Dundas}},
  \bibinfo{author}{\bibfnamefont{E.~J.} \bibnamefont{McEniry}},
  \bibnamefont{and} \bibinfo{author}{\bibfnamefont{T.~T.} \bibnamefont{N.}},
  \bibinfo{journal}{Nat. Nanotechnol.}, \textbf{\bibinfo{volume}{4}}, \bibinfo{paes}{99}(\bibinfo{year}{2009}).

\bibitem[{\citenamefont{L\"{u} et~al.}(2010)\citenamefont{L\"{u}, Brandbyge, and
  Hedeg\r{a}rd}}]{hedegaard}
\bibinfo{author}{\bibfnamefont{J.-T.} \bibnamefont{L\"{u}}},
  \bibinfo{author}{\bibfnamefont{M.}~\bibnamefont{Brandbyge}},
  \bibnamefont{and}
  \bibinfo{author}{\bibfnamefont{P.}~\bibnamefont{Hedeg\r{a}rd}},
  \bibinfo{journal}{Nano Lett.} \textbf{\bibinfo{volume}{10}},
  \bibinfo{pages}{1657} (\bibinfo{year}{2010}).

\bibitem[{\citenamefont{Enkovaara et~al.}(2010)\citenamefont{Enkovaara,
  Rostgaard, Mortensen, Chen, Du{\l}ak, Ferrighi, Gavnholt, Glinsvad, Haikola,
  Hansen et~al.}}]{GPAW_review}
\bibinfo{author}{\bibfnamefont{J.}~\bibnamefont{Enkovaara}},
  \bibinfo{author}{\bibfnamefont{C.}~\bibnamefont{Rostgaard}},
  \bibinfo{author}{\bibfnamefont{J.~J.} \bibnamefont{Mortensen}},
  \bibinfo{author}{\bibfnamefont{J.}~\bibnamefont{Chen}},
  \bibinfo{author}{\bibfnamefont{M.}~\bibnamefont{Du{\l}ak}},
  \bibinfo{author}{\bibfnamefont{L.}~\bibnamefont{Ferrighi}},
  \bibinfo{author}{\bibfnamefont{J.}~\bibnamefont{Gavnholt}},
  \bibinfo{author}{\bibfnamefont{C.}~\bibnamefont{Glinsvad}},
  \bibinfo{author}{\bibfnamefont{V.}~\bibnamefont{Haikola}},
  \bibinfo{author}{\bibfnamefont{H.~A.} \bibnamefont{Hansen}},
  \bibnamefont{et~al.}, \bibinfo{journal}{Journal of Physics: Condensed Matter}
  \textbf{\bibinfo{volume}{22}}, \bibinfo{pages}{253202}
  (\bibinfo{year}{2010}).
  
\bibitem[{\citenamefont{Larsen et~al.}(2009)\citenamefont{Larsen, Vanin,
  Mortensen, Thygesen, and Jacobsen}}]{gpaw-lcao}
\bibinfo{author}{\bibfnamefont{A.~H.} \bibnamefont{Larsen}},
  \bibinfo{author}{\bibfnamefont{M.}~\bibnamefont{Vanin}},
  \bibinfo{author}{\bibfnamefont{J.~J.} \bibnamefont{Mortensen}},
  \bibinfo{author}{\bibfnamefont{K.~S.} \bibnamefont{Thygesen}},
  \bibnamefont{and} \bibinfo{author}{\bibfnamefont{K.~W.}
  \bibnamefont{Jacobsen}}, \bibinfo{journal}{Phys. Rev. B}
  \textbf{\bibinfo{volume}{80}}, \bibinfo{pages}{195112}
  (\bibinfo{year}{2009}).

\bibitem[{\citenamefont{Taylor et~al.}(2001{\natexlab{a}})\citenamefont{Taylor,
  Guo, and Wang}}]{taylor}
\bibinfo{author}{\bibfnamefont{J.}~\bibnamefont{Taylor}},
  \bibinfo{author}{\bibfnamefont{H.}~\bibnamefont{Guo}}, \bibnamefont{and}
  \bibinfo{author}{\bibfnamefont{J.}~\bibnamefont{Wang}},
  \bibinfo{journal}{Phys. Rev. B} \textbf{\bibinfo{volume}{63}},
  \bibinfo{pages}{245407} (\bibinfo{year}{2001}{\natexlab{a}}).

\bibitem[{\citenamefont{Brandbyge et~al.}(2002)\citenamefont{Brandbyge, Mozos,
  Ordej\'on, Taylor, and Stokbro}}]{BrandbyNEGF}
\bibinfo{author}{\bibfnamefont{M.}~\bibnamefont{Brandbyge}},
  \bibinfo{author}{\bibfnamefont{J.-L.} \bibnamefont{Mozos}},
  \bibinfo{author}{\bibfnamefont{P.}~\bibnamefont{Ordej\'on}},
  \bibinfo{author}{\bibfnamefont{J.}~\bibnamefont{Taylor}}, \bibnamefont{and}
  \bibinfo{author}{\bibfnamefont{K.}~\bibnamefont{Stokbro}},
  \bibinfo{journal}{Phys. Rev. B} \textbf{\bibinfo{volume}{65}},
  \bibinfo{pages}{165401} (\bibinfo{year}{2002}).

\bibitem[{\citenamefont{Xue et~al.}(2002)\citenamefont{Xue, Datta, and
  Ratner}}]{xue}
\bibinfo{author}{\bibfnamefont{Y.}~\bibnamefont{Xue}},
  \bibinfo{author}{\bibfnamefont{S.}~\bibnamefont{Datta}}, \bibnamefont{and}
  \bibinfo{author}{\bibfnamefont{M.~A.} \bibnamefont{Ratner}},
  \bibinfo{journal}{Chemical Physics} \textbf{\bibinfo{volume}{281}},
  \bibinfo{pages}{151 } (\bibinfo{year}{2002}).

\bibitem[{\citenamefont{Guinea et~al.}(1983)\citenamefont{Guinea, Tejedor,
  Flores, and Louis}}]{guinea}
\bibinfo{author}{\bibfnamefont{F.}~\bibnamefont{Guinea}},
  \bibinfo{author}{\bibfnamefont{C.}~\bibnamefont{Tejedor}},
  \bibinfo{author}{\bibfnamefont{F.}~\bibnamefont{Flores}}, \bibnamefont{and}
  \bibinfo{author}{\bibfnamefont{E.}~\bibnamefont{Louis}},
  \bibinfo{journal}{Phys. Rev. B} \textbf{\bibinfo{volume}{28}},
  \bibinfo{pages}{4397} (\bibinfo{year}{1983}).

\bibitem{jauho}
See, for example, H. Haug and A.-P. Jauho,
\newblock \emph{Quantum Kinetics in Transport and Optics of Semiconductors} (Springer-Verlag, New York, 1998).

\bibitem[{\citenamefont{Kiejna et~al.}(2006)\citenamefont{Kiejna, Kresse,
  Rogal, De~Sarkar, Reuter, and Scheffler}}]{Kresse}
\bibinfo{author}{\bibfnamefont{A.}~\bibnamefont{Kiejna}},
  \bibinfo{author}{\bibfnamefont{G.}~\bibnamefont{Kresse}},
  \bibinfo{author}{\bibfnamefont{J.}~\bibnamefont{Rogal}},
  \bibinfo{author}{\bibfnamefont{A.}~\bibnamefont{De~Sarkar}},
  \bibinfo{author}{\bibfnamefont{K.}~\bibnamefont{Reuter}}, \bibnamefont{and}
  \bibinfo{author}{\bibfnamefont{M.}~\bibnamefont{Scheffler}},
  \bibinfo{journal}{Phys. Rev. B} \textbf{\bibinfo{volume}{73}},
  \bibinfo{pages}{035404} (\bibinfo{year}{2006}).

\bibitem[{\citenamefont{Meir and Wingreen}(1992)}]{meir}
\bibinfo{author}{\bibfnamefont{Y.}~\bibnamefont{Meir}} \bibnamefont{and}
  \bibinfo{author}{\bibfnamefont{N.~S.} \bibnamefont{Wingreen}},
  \bibinfo{journal}{Phys. Rev. Lett.} \textbf{\bibinfo{volume}{68}},
  \bibinfo{pages}{2512} (\bibinfo{year}{1992}).
 
\bibitem[{\citenamefont{Thygesen}(2006)}]{nonorthogonal}
\bibinfo{author}{\bibfnamefont{K.~S.} \bibnamefont{Thygesen}},
  \bibinfo{journal}{Phys. Rev. B} \textbf{\bibinfo{volume}{73}},
  \bibinfo{pages}{035309} (\bibinfo{year}{2006}).
  
\bibitem[{\citenamefont{Taylor et~al.}(2001{\natexlab{b}})\citenamefont{Taylor,
  Guo, and Wang}}]{TaylorNEGF}
\bibinfo{author}{\bibfnamefont{J.}~\bibnamefont{Taylor}},
  \bibinfo{author}{\bibfnamefont{H.}~\bibnamefont{Guo}}, \bibnamefont{and}
  \bibinfo{author}{\bibfnamefont{J.}~\bibnamefont{Wang}},
  \bibinfo{journal}{Phys. Rev. B} \textbf{\bibinfo{volume}{63}},
  \bibinfo{pages}{245407} (\bibinfo{year}{2001}{\natexlab{b}}).

\bibitem[{\citenamefont{Li}(2008)}]{LiruiThesis}
\bibinfo{author}{\bibfnamefont{R.}~\bibnamefont{Li}}, \bibinfo{journal}{Phd
  Thesis, School of Electronics and Computer Science, Peking University}
  (\bibinfo{year}{2008}).

\bibitem[{\citenamefont{Patterson}(1968)}]{patterson}
\bibinfo{author}{\bibfnamefont{T.~N.~L.} \bibnamefont{Patterson}},
  \bibinfo{journal}{Math. Comput.} \textbf{\bibinfo{volume}{22}},
  \bibinfo{pages}{847} (\bibinfo{year}{1968}).

\bibitem[{\citenamefont{Zekan et~al.}(2008)\citenamefont{Zekan, Shimin, Rui,
  Ziyong, Xingyu, and Zengquan}}]{SparseMatrixQian}
\bibinfo{author}{\bibfnamefont{Q.}~\bibnamefont{Zekan}},
  \bibinfo{author}{\bibfnamefont{H.}~\bibnamefont{Shimin}},
  \bibinfo{author}{\bibfnamefont{L.}~\bibnamefont{Rui}},
  \bibinfo{author}{\bibfnamefont{S.}~\bibnamefont{Ziyong}},
  \bibinfo{author}{\bibfnamefont{Z.}~\bibnamefont{Xingyu}}, \bibnamefont{and}
  \bibinfo{author}{\bibfnamefont{X.}~\bibnamefont{Zengquan}},
  \bibinfo{journal}{J. Comput. Theor. Nanosci.} \textbf{\bibinfo{volume}{5}},
  \bibinfo{pages}{671} (\bibinfo{year}{2008}).

\bibitem[{\citenamefont{Ozaki et~al.}(2010)\citenamefont{Ozaki, Nishio, and
  Kino}}]{JapanNEGF}
\bibinfo{author}{\bibfnamefont{T.}~\bibnamefont{Ozaki}},
  \bibinfo{author}{\bibfnamefont{K.}~\bibnamefont{Nishio}}, \bibnamefont{and}
  \bibinfo{author}{\bibfnamefont{H.}~\bibnamefont{Kino}},
  \bibinfo{journal}{Phys. Rev. B} \textbf{\bibinfo{volume}{81}},
  \bibinfo{pages}{035116} (\bibinfo{year}{2010}).

\bibitem[{\citenamefont{Di~Ventra et~al.}(2004)\citenamefont{Di~Ventra, Chen,
  and Todorov}}]{Non_conservative_force}
\bibinfo{author}{\bibfnamefont{M.}~\bibnamefont{Di~Ventra}},
  \bibinfo{author}{\bibfnamefont{Y.-C.} \bibnamefont{Chen}}, \bibnamefont{and}
  \bibinfo{author}{\bibfnamefont{T.~N.} \bibnamefont{Todorov}},
  \bibinfo{journal}{Phys. Rev. Lett.} \textbf{\bibinfo{volume}{92}},
  \bibinfo{pages}{176803} (\bibinfo{year}{2004}).

\bibitem[{\citenamefont{B.Oostinga et~al.}(2008)\citenamefont{B.Oostinga,
  B.Heersche, Liu, F.Morpurgo, and M.K.Vandersypen}}]{Gate_BilayerGNR}
\bibinfo{author}{\bibfnamefont{J.}~\bibnamefont{B.Oostinga}},
  \bibinfo{author}{\bibfnamefont{H.}~\bibnamefont{B.Heersche}},
  \bibinfo{author}{\bibfnamefont{X.}~\bibnamefont{Liu}},
  \bibinfo{author}{\bibfnamefont{A.}~\bibnamefont{F.Morpurgo}},
  \bibnamefont{and}
  \bibinfo{author}{\bibfnamefont{L.}~\bibnamefont{M.K.Vandersypen}},
  \bibinfo{journal}{Nature Mater.},\textbf{\bibinfo{volume}{7}}, \bibinfo{pages}{151}
  (\bibinfo{year}{2008}). 

\bibitem[{\citenamefont{M.F.Craciun et~al.}(2009)\citenamefont{M.F.Craciun,
  S.Russo, M.Yamamoto, B.Oostinga, F.Morpurgo, and
  S.Tarucha}}]{Gate_TrilayerGNR}
\bibinfo{author}{\bibnamefont{M.F.Craciun}},
  \bibinfo{author}{\bibnamefont{S.Russo}},
  \bibinfo{author}{\bibnamefont{M.Yamamoto}},
  \bibinfo{author}{\bibfnamefont{J.}~\bibnamefont{B.Oostinga}},
  \bibinfo{author}{\bibfnamefont{A.}~\bibnamefont{F.Morpurgo}},
  \bibnamefont{and} \bibinfo{author}{\bibnamefont{S.Tarucha}},
  \bibinfo{journal}{Nat. Nanotechnol.}, \textbf{\bibinfo{volume}{4}},\bibinfo{pages}{383}
  (\bibinfo{year}{2009}).

\bibitem[{\citenamefont{Kubatkin et~al.}(2003)\citenamefont{Kubatkin, Danilov,
  Hjort, Cornil, Bredas, Stuhr-Hansen, Hedegard, and Bjornholm}}]{kubatkin}
\bibinfo{author}{\bibfnamefont{S.}~\bibnamefont{Kubatkin}},
  \bibinfo{author}{\bibfnamefont{A.}~\bibnamefont{Danilov}},
  \bibinfo{author}{\bibfnamefont{M.}~\bibnamefont{Hjort}},
  \bibinfo{author}{\bibfnamefont{J.}~\bibnamefont{Cornil}},
  \bibinfo{author}{\bibfnamefont{J.-L.} \bibnamefont{Bredas}},
  \bibinfo{author}{\bibfnamefont{N.}~\bibnamefont{Stuhr-Hansen}},
  \bibinfo{author}{\bibfnamefont{P.}~\bibnamefont{Hedegard}}, \bibnamefont{and}
  \bibinfo{author}{\bibfnamefont{T.}~\bibnamefont{Bjornholm}},
  \bibinfo{journal}{Nature},\textbf{\bibinfo{volume}{425}},\bibinfo{pages}{698} (\bibinfo{year}{2003}).

\bibitem[{\citenamefont{Hyunwook et~al.}(2009)\citenamefont{Hyunwook,
  Youngsang, Hee, Heejun, A., and Takhee}}]{Gate_Mark_Reed}
\bibinfo{author}{\bibfnamefont{S.}~\bibnamefont{Hyunwook}},
  \bibinfo{author}{\bibfnamefont{K.}~\bibnamefont{Youngsang}},
  \bibinfo{author}{\bibfnamefont{J.~Y.} \bibnamefont{Hee}},
  \bibinfo{author}{\bibfnamefont{J.}~\bibnamefont{Heejun}},
  \bibinfo{author}{\bibfnamefont{R.~M.} \bibnamefont{A.}}, \bibnamefont{and}
  \bibinfo{author}{\bibfnamefont{L.}~\bibnamefont{Takhee}},
  \bibinfo{journal}{Nature},\textbf{\bibinfo{volume}{462}}, \bibinfo{pages}{1039} (\bibinfo{year}{2009}).

\bibitem[{\citenamefont{Datta et~al.}(2009)\citenamefont{Datta, Strachan, and
  Johnson}}]{Datta_gate}
\bibinfo{author}{\bibfnamefont{S.~S.} \bibnamefont{Datta}},
  \bibinfo{author}{\bibfnamefont{D.~R.} \bibnamefont{Strachan}},\bibnamefont{and}
  \bibinfo{author}{\bibfnamefont{A.~T.~Charlie} \bibnamefont{Johnson}}, 
  \bibinfo{journal}{Phys. Rev. B}
  \textbf{\bibinfo{volume}{79}}, \bibinfo{pages}{205404}
  (\bibinfo{year}{2009}).

\bibitem[{\citenamefont{Strange et~al.}(2010)\citenamefont{Strange,
  Lopez-Acevedo, and H\"akkinen}}]{hakkinen}
\bibinfo{author}{\bibfnamefont{M.}~\bibnamefont{Strange}},
  \bibinfo{author}{\bibfnamefont{O.}~\bibnamefont{Lopez-Acevedo}},
  \bibnamefont{and}
  \bibinfo{author}{\bibfnamefont{H.}~\bibnamefont{H\"akkinen}},
  \bibinfo{journal}{J. Phys. Chem. Lett.} \textbf{\bibinfo{volume}{1}},
  \bibinfo{pages}{1528} (\bibinfo{year}{2010}).

\bibitem[{\citenamefont{Jadzinsky et~al.}(2007)\citenamefont{Jadzinsky, Calero,
  Ackerson, Bushnell, and Kornberg}}]{jadzinsky}
\bibinfo{author}{\bibfnamefont{P.~D.} \bibnamefont{Jadzinsky}},
  \bibinfo{author}{\bibfnamefont{G.}~\bibnamefont{Calero}},
  \bibinfo{author}{\bibfnamefont{C.~J.} \bibnamefont{Ackerson}},
  \bibinfo{author}{\bibfnamefont{D.~A.} \bibnamefont{Bushnell}},
  \bibnamefont{and} \bibinfo{author}{\bibfnamefont{R.~D.}
  \bibnamefont{Kornberg}}, \bibinfo{journal}{Science}
  \textbf{\bibinfo{volume}{318}}, \bibinfo{pages}{430} (\bibinfo{year}{2007}).

\bibitem[{\citenamefont{Wang et~al.}(2009)\citenamefont{Wang, Chi, Hush,
  Reimers, Zhang, and Ulstrup}}]{ulstrup}
\bibinfo{author}{\bibfnamefont{Y.}~\bibnamefont{Wang}},
  \bibinfo{author}{\bibfnamefont{Q.}~\bibnamefont{Chi}},
  \bibinfo{author}{\bibfnamefont{N.~S.} \bibnamefont{Hush}},
  \bibinfo{author}{\bibfnamefont{J.~R.} \bibnamefont{Reimers}},
  \bibinfo{author}{\bibfnamefont{J.}~\bibnamefont{Zhang}}, \bibnamefont{and}
  \bibinfo{author}{\bibfnamefont{J.}~\bibnamefont{Ulstrup}},
  \bibinfo{journal}{J. Phys. Chem. C} \textbf{\bibinfo{volume}{113}},
  \bibinfo{pages}{19601} (\bibinfo{year}{2009}).

\bibitem[{\citenamefont{Markussen
  et~al.}(2010{\natexlab{b}})\citenamefont{Markussen, Schiotz, and
  Thygesen}}]{TroelsQI}
\bibinfo{author}{\bibfnamefont{T.}~\bibnamefont{Markussen}},
  \bibinfo{author}{\bibfnamefont{J.}~\bibnamefont{Schiotz}}, \bibnamefont{and}
  \bibinfo{author}{\bibfnamefont{K.~S.} \bibnamefont{Thygesen}},
  \bibinfo{journal}{The Journal of Chemical Physics}
  \textbf{\bibinfo{volume}{132}}, \bibinfo{eid}{224104}(\bibinfo{year}{2010}{\natexlab{b}}).

\bibitem[{\citenamefont{Markussen
  et~al.}(2010{\natexlab{c}})\citenamefont{Markussen, Stadler, and
  Thygesen}}]{troels2}
\bibinfo{author}{\bibfnamefont{T.}~\bibnamefont{Markussen}},
  \bibinfo{author}{\bibfnamefont{R.}~\bibnamefont{Stadler}}, \bibnamefont{and}
  \bibinfo{author}{\bibfnamefont{K.~S.} \bibnamefont{Thygesen}},
  \bibinfo{journal}{Nano Lett.} \textbf{\bibinfo{volume}{10}},
  \bibinfo{pages}{4260} (\bibinfo{year}{2010}{\natexlab{c}}).

\bibitem[{\citenamefont{Areshkin and Nikoli\ifmmode~\acute{c}\else
  \'{c}\fi{}}(2009)}]{Branislav}
\bibinfo{author}{\bibfnamefont{D.~A.} \bibnamefont{Areshkin}} \bibnamefont{and}
  \bibinfo{author}{\bibfnamefont{B.~K.}
  \bibnamefont{Nikoli\ifmmode~\acute{c}\else \'{c}\fi{}}},
  \bibinfo{journal}{Phys. Rev. B} \textbf{\bibinfo{volume}{79}},
  \bibinfo{pages}{205430} (\bibinfo{year}{2009}).

\bibitem[{\citenamefont{Fujita et~al.}(1996)\citenamefont{Fujita, Wakabayashi,
  Nakada, and Kusakabe}}]{spin_polarized_edge_state}
\bibinfo{author}{\bibfnamefont{M.}~\bibnamefont{Fujita}},
  \bibinfo{author}{\bibfnamefont{K.}~\bibnamefont{Wakabayashi}},
  \bibinfo{author}{\bibfnamefont{K.}~\bibnamefont{Nakada}}, \bibnamefont{and}
  \bibinfo{author}{\bibfnamefont{K.}~\bibnamefont{Kusakabe}},
  \bibinfo{journal}{Journal of the Physical Society of Japan}
  \textbf{\bibinfo{volume}{65}}, \bibinfo{pages}{1920} (\bibinfo{year}{1996}).

\bibitem[{\citenamefont{Zhang et~al.}(2005)\citenamefont{Zhang, Hou, Li, Qian,
  Han, Shen, Zhao, and Xue}}]{zhangjiaxing}
\bibinfo{author}{\bibfnamefont{J.}~\bibnamefont{Zhang}},
  \bibinfo{author}{\bibfnamefont{S.}~\bibnamefont{Hou}},
  \bibinfo{author}{\bibfnamefont{R.}~\bibnamefont{Li}},
  \bibinfo{author}{\bibfnamefont{Z.}~\bibnamefont{Qian}},
  \bibinfo{author}{\bibfnamefont{R.}~\bibnamefont{Han}},
  \bibinfo{author}{\bibfnamefont{Z.}~\bibnamefont{Shen}},
  \bibinfo{author}{\bibfnamefont{X.}~\bibnamefont{Zhao}}, \bibnamefont{and}
  \bibinfo{author}{\bibfnamefont{Z.}~\bibnamefont{Xue}},
  \bibinfo{journal}{Nanotechnology} \textbf{\bibinfo{volume}{16}},
  \bibinfo{pages}{3057} (\bibinfo{year}{2005}).
  \end{thebibliography}
\end{document}